\documentclass[12pt]{article} 
\usepackage[sectionbib]{natbib}
\usepackage{array,epsfig,fancyheadings,rotating}
\usepackage[]{hyperref}  
\usepackage{sectsty, secdot}
\sectionfont{\fontsize{12}{14pt plus.8pt minus .6pt}\selectfont}
\renewcommand{\theequation}{\thesection\arabic{equation}}
\subsectionfont{\fontsize{12}{14pt plus.8pt minus .6pt}\selectfont}

\usepackage{amsmath}
\usepackage{amssymb}
\usepackage{amsfonts}
\usepackage{multirow}
\usepackage{amsthm}

\newtheorem{theorem}{Theorem}
\newtheorem{lemma}{Lemma}

\theoremstyle{definition}

\begin{document}


\renewcommand{\baselinestretch}{1.1}

\markright{ \hbox{\footnotesize\rm 
}\hfill\\[-13pt]
\hbox{\footnotesize\rm
}\hfill }

\markboth{\hfill{\footnotesize\rm FIRSTNAME1 LASTNAME1 AND FIRSTNAME2 LASTNAME2} \hfill}
{\hfill {\footnotesize\rm FILL IN A SHORT RUNNING TITLE} \hfill}

\renewcommand{\thefootnote}{}
$\ $\par


\fontsize{12}{14pt plus.8pt minus .6pt}\selectfont \vspace{0.8pc}
\centerline{\large\bf ROBUST OPTIMAL DESIGNS WHEN MISSING}
\vspace{2pt} 
\centerline{\large\bf  DATA HAPPEN AT RANDOM}
\vspace{.4cm} 
\centerline{Rui Hu, Ion Bica, Zhichun Zhai} 
\vspace{.4cm} 
\centerline{\it MacEwan University}
 \vspace{.55cm} \fontsize{9}{11.5pt plus.8pt minus.6pt}\selectfont


\begin{quotation}
\noindent {\it Abstract:}
In this article, we investigate the robust optimal design problem for the prediction of response when the fitted regression models are only approximately specified, and observations might be missing completely at random. The intuitive idea is as follows: We assume that data are missing at random, and the complete case analysis is applied. To account for the occurrence of missing data, the design criterion we choose is the mean, for the missing indicator, of the averaged (over the design space) mean squared errors of the predictions. To describe the uncertainty in the specification of the real underlying model, we impose a neighborhood structure on the deterministic part of the regression response and maximize,  analytically, the \textbf{M}ean of the averaged \textbf{M}ean squared \textbf{P}rediction \textbf{E}rrors  (MMPE), over the entire neighborhood. The maximized MMPE is the ``worst'' loss in the neighborhood of the fitted regression model. Minimizing the maximum MMPE over the class of designs, we obtain robust ``minimax'' designs. The robust designs constructed afford protection from increases in prediction errors resulting from model misspecifications.

\vspace{9pt}
\noindent {\it Key words and phrases:}
 Minimax,  missing completely at random, missing observations,  optimal designs, robustness.
\par
\end{quotation}\par

\def\thefigure{\arabic{figure}}
\def\thetable{\arabic{table}}

\renewcommand{\theequation}{\thesection.\arabic{equation}}

\fontsize{12}{14pt plus.8pt minus .6pt}\selectfont

\section{Introduction}

Data missing is a common problem in many practices. As pointed out in \cite{Schafer1997} improper handling of missing values may reduce the power of the study, increase the variability of estimation, cause bias in estimates, and result in misleading conclusions. Missing data mechanisms roughly can be categorized into missing at random (MAR) and missing not at random (MNAR). Data missing at random (MAR), as defined in \cite{Rubin1976}, means that the missingness depends only on the observed data, and not on the components that are missing.

A large amount of literature proposed various methods for handling data missing at random. The classical and popular methods include complete case analysis, the hot deck imputation proposed in \cite{OS1980}, the last observation carried forward developed by \cite{SZ2003}, and so on. Among them, the simplest and the most direct method is complete case analysis, i.e., removal of all cases with missing data. This method is  commonly used  for data missing completely at random (MCAR), which is a special case of MAR. The detailed explanation is available in \cite{Little1992} and \cite{LR2019}. 

 A few studies are focusing on handling the missing data problem at the design stage. \cite{OTB2008} studied D-optimal designs for linear mixed models where dropout is encountered in the longitudinal data. \cite{AG2010} assessed the robustness of subset designs, a class of response surface designs, against missing observations for the efficiency of parameter estimation. \cite{ISW2002}  proposed a method for constructing efficient designs by assuming varying-probabilities of realizing responses. \cite{LBM2018}  developed a new approach for proposing optimal designs, which include the method proposed in \cite{ISW2002} as a special case.  
 
Motivated by the work in \cite{LBM2018}, in this paper, we will study the construction of optimal designs that are robust in that we allow imprecision in the specification of the response. The development of robust designs when missing data might occur has aroused attention.  \cite{ISW2002}  discussed the construction of optimal designs when the response probability function belongs to a known set of plausible functions. \cite{ISW2004}    investigated the sensitivity of their optimal designs to nominal values assigned to the parameters in the model.  In our framework, the experimenter assumes - perhaps erroneously - that the mean conditional response is given by a  known function of unknown parameters and known regressors.  We will propose  optimal designs that are robust to the potential model misspecification.

Let $\mathbf{x}_i, i=1,..., n$, be a set of regressors for the $i$-th subject in the experiment and $\mathcal{S}=\{\mathbf{x}_1, ..., \mathbf{x}_N\}$ be the design space.  A design is a specification of weights for the points in the design space $\boldsymbol{\xi}=(\xi_1, ..., \xi_N)$ where $\xi_i=n_i/n$ such that $\sum_{i=1}^Nn_i=n$. Then  $n_i$ observations are made at the covariate $\mathbf{x}_i$. For convenience, we denote $diag(n\boldsymbol{\xi})$ as $\mathbf{D}_{\boldsymbol{\xi}}$. 
Since the responses may be missing,  the missing indicator is useful for the analysis and can be defined as following
\begin{eqnarray}
m_{ij}=\left\{\begin{array}{ll}
1     & \hbox{if the $j$th observation at $\mathbf{x}_i$ is not missing}\\
 0   &  \hbox{if the $j$th observation at $\mathbf{x}_i$ is  missing}
\end{array}\right.
\label{MARE11}
\end{eqnarray}
with $i=1, ..., N$ and $j\leq n_i$.
For predictors not selected in the design, their missing data indicators are 0. Denote  $\mathbf{M}=(m_{11},...,m_{1n_1}, ..., m_{N1},..., m_{Nn_N})^T.$  Similar to \cite{LBM2018}, we consider the scenario that the responses are missing completely at random  (MCAR). The probability of response missing at $\mathbf{x}_i$ is $1-p(m_{ij}=1|\mathbf{x}_i, \boldsymbol{\gamma})$ where the probabilities of being missing  $p(m_{ij}=1|\mathbf{x}_i, \boldsymbol{\gamma})$ is assumed to only depend on $\mathbf{x}_i$ and nuisance parameters $\boldsymbol{\gamma}$.
Another reasonable assumption about the missing indicator is that $\sum_{i=1}^N\sum_{j=1}^{n\xi_i}(1-m_{ij})=O_p(1)$. That is, the number of missing data,  in probability,  increases only at the speed of $O(1)$.

Assume that  the experimenter fits a  regression model to the data of the following form
\begin{eqnarray}
Y=f(\mathbf{x};\boldsymbol{\beta})+{\varepsilon}
\label{MARE1}
\end{eqnarray}
where 
$\mathbf{x}=(x_1,...,x_p)$ and $\boldsymbol{\beta}$ is the regression coefficients vector. The random errors  $(\varepsilon)$ are independent and identically distributed with mean of 0 and variance of $\sigma^2$. 
 
For robustness, one anticipates that the model (\ref{MARE1}) fitted by the experimenter is not necessarily the true one. The deviation from the model assumption may lead to biased predict of the response variable (see, for example, \cite{BD1959}, \cite{FTK1989}, \cite{XY2011}, \cite{MZWF2017}). Therefore, in this paper, we will propose the robust designs that optimize the precision of prediction over a small neighborhood to which the true model might belong.

Let
\begin{eqnarray}
n^{-1/2}\psi(\mathbf{x};\boldsymbol{\beta})=E[Y|\mathbf{x}]-f(\mathbf{x};\boldsymbol{\beta}),
\label{MARE15}
\end{eqnarray}
so that the exact but only approximately specified model is 
\begin{eqnarray}
Y=f(\mathbf{x};\boldsymbol{\beta})+n^{-1/2}\psi(\mathbf{x};\boldsymbol{\beta})+\varepsilon.
\label{MARE2}
\end{eqnarray}
The ``true'' $\boldsymbol{\beta}$ is defined through minimizing the integrated squared discrepancy as follows
\begin{eqnarray}
\boldsymbol{\beta}=\arg\min_{\boldsymbol{\theta}} \int_{\mathcal{S}}\psi^2(\mathbf{x};\boldsymbol{\theta}) d\mathbf{x}.
\label{MARE3}
\end{eqnarray}
For convenience, let
$$\mathbf{z}_i(\boldsymbol{\beta})= \frac{\partial{f}(\mathbf{x}_i;\boldsymbol{\beta})}{\partial \boldsymbol{\beta}}$$
and 
$$\mathbf{Z}(\boldsymbol{\beta})=\left(\begin{array}{c}
     \mathbf{z}^T_1(\boldsymbol{\beta})  \\
     ...\\
     \mathbf{z}^T_N(\boldsymbol{\beta}) 
\end{array} \right)_{N\times p}.$$
Then, due to (\ref{MARE3}), a constraint on $\Psi(\boldsymbol{\beta})= (\psi(\mathbf{x}_1;\boldsymbol{\beta}), ...,\psi(\mathbf{x}_N;\boldsymbol{\beta}))^T $ is
\begin{eqnarray}
\Psi^T(\boldsymbol{\beta})\mathbf{Z}(\boldsymbol{\beta})=\mathbf{0}.
 \nonumber
\end{eqnarray}

We consider a small neighborhood of $\Psi(\boldsymbol{\beta})$ as follows
\begin{eqnarray}
\mathcal{F}=\left\{\Psi(\boldsymbol{\beta}):\quad  \Psi^T(\boldsymbol{\beta})\mathbf{Z}(\boldsymbol{\beta})=\mathbf{0},\quad  \|\Psi(\boldsymbol{\beta}) \|^2\leq \eta^2<+\infty \right\}.
\label{MARE4}
\end{eqnarray}
That is, the true model is assumed to be in the small neighborhood of  (\ref{MARE1}): $\{F(\boldsymbol{\beta})+n^{-1/2}\Psi(\boldsymbol{\beta}):\Psi(\boldsymbol{\beta}) \in \mathcal{F}\}$ with  $F(\boldsymbol{\beta})=(f(\mathbf{x}_1;\boldsymbol{\beta}),..., f(\mathbf{x}_N;\boldsymbol{\beta}))^T$. In the following, without confusion, we will refer to this neighbourhood of the model (\ref{MARE1}) also as $\mathcal{F}$.

Experimenters often do not realize the misspecification of the fitted model (\ref{MARE1}).  So they may still estimate the regression parameters by the maximum likelihood estimate $\hat{\boldsymbol{\beta}}$. This estimate, however, will lead to an incorrect prediction of the response values. It is our intention to propose optimal designs that minimize the ``worst'' (i.e., the largest over $\mathcal{F}$) \textbf{A}veraged \textbf{M}ean \textbf{S}quared \textbf{E}rrors (AMSE) of the predicted values over $\mathcal{S}$ where 
$$AMSE=\frac{1}{N}\sum_{i=1}^N  E_{\hat{\boldsymbol{\beta}}}[ f(\mathbf{x}_i,\hat{\boldsymbol{\beta}})-E(Y|\mathbf{x}_i)]^2. $$
Furthermore, to account for the occurrence of missing data, we take the mean of AMSE with respect to the missing indicators (\ref{MARE11}). Thus the $MMPE(\psi,{\boldsymbol{\xi}})$ defined below will be used as a measure of loss:
\begin{eqnarray}
MMPE(\psi,{\boldsymbol{\xi}})
&=& E_{\mathbf{M}}\left[\frac{1}{N}\sum_{i=1}^N  E_{\hat{\boldsymbol{\beta}}}[ f(\mathbf{x}_i,\hat{\boldsymbol{\beta}})-E(Y|\mathbf{x}_i)]^2\right]\nonumber\\
&=& \frac{1}{N}\sum_{i=1}^N  E_{\hat{\boldsymbol{\beta}},\mathbf{M}}[ f(\mathbf{x}_i;\hat{\boldsymbol{\beta}})-E(Y|\mathbf{x}_i)]^2.
\label{MARE8}
\end{eqnarray}

A problem that arises immediately is that when the regression model is nonlinear $MMPE(\psi,{\boldsymbol{\xi}})$ in  (\ref{MARE8}) depends on the unknown values of the parameters $\boldsymbol{\beta}$.  There are various methods for handling this problem. One is by constructing a \textquotedblleft locally optimal\textquotedblright\ design -- one that is optimal only at a
particular value $\boldsymbol{\beta}_{0}$ of the parameter. 
To allow for uncertainty about the parameter values, an approach is to first maximize the loss function over a neighborhood of a local parameter $\boldsymbol{\beta}_{0}$ and then minimize the maximized loss function over the class of designs. 
Sequential strategy is also applied  to address the parameter dependence issue (see, for example, \cite{SW2002}, \cite{HW2017}, \cite{Hu2018}). In this approach, the estimates are evaluated using the available data, and subsequent observations are made at new design points minimizing the loss function, evaluated at the current estimates. In this paper, we will apply Bayesian methods, which are discussed and widely used in literature such as \cite{DN1997}, \cite{KW2014}. By applying the Bayesian method, the loss function is averaged with respect to an appropriate prior distribution on the parameters before being minimized.

 In  section \ref{sec-Max}, we find the ``worst'' value of $MMPE(\psi, \boldsymbol{\xi})$ over $\mathcal{F}$, theoretically. The linear and nonlinear regression models are investigated separately. For linear models, AMSE can be found explicitly. For nonlinear models, an asymptotic approximation of AMSE has been derived. Since it is impossible to calculate $E_{\mathbf{M}}[AMSE]$ directly, we will find an approximation of MMSE by taking the expectation for the missing indicator on the first term and the second term of the Taylor expansion of AMSE. In Section \ref{Sec numeric}, we  apply the generic algorithm  to look for the minimax robust optimal designs by minimizing the largest $MMPE(\psi, \boldsymbol{\xi})$ regarding the design ${\boldsymbol{\xi}}$. The development of robust designs concerns linear regression and nonlinear regression examples.  Computing code, written in Matlab, to duplicate these examples is available from the authors. All the detailed derivations are in the Appendices A-E.

\section{Maximum of $MMPE(\psi, \boldsymbol{\xi})$ over $\mathcal{F}$} \label{sec-Max}
We will decompose $MMPE(\psi, \boldsymbol{\xi})$ into four terms. This decomposition will have a more explicit form for the multiple linear regression model, which will lead us to define the loss function as the Taylor approximate of it.
\begin{lemma}\label{Throem decompse of loss}
 The $MMPE(\psi, \boldsymbol{\xi})$ defined in (\ref{MARE8}) can be decomposed as follows
\begin{eqnarray}
MMPE(\psi, \boldsymbol{\xi})
=MB(\boldsymbol{\xi})+MV(\boldsymbol{\xi})-\frac{2}{N} E_{\mathbf{M}}[B^T(\boldsymbol{\xi})] \frac{\Psi(\boldsymbol{\beta})}{\sqrt{n}}  
+\frac{1}{Nn}\|\Psi(\boldsymbol{\beta})\|^2
\label{MARE5}
\end{eqnarray}
where the bias vector $B(\boldsymbol{\xi})$ is
\begin{eqnarray}
B(\boldsymbol{\xi})=\left(E_{\hat{\boldsymbol{\beta}}}[f(\boldsymbol{x}_i; \hat{\boldsymbol{\beta}})]-f(\boldsymbol{x}_i;\boldsymbol{\beta}) \right)_{i=1}^N
\nonumber
\end{eqnarray}
with the mean bias $MB(\boldsymbol{\xi})$, and the mean  variance $MV(\boldsymbol{\xi})$ being
\begin{eqnarray}
MB(\boldsymbol{\xi})&=&\frac{1}{N} E_{\mathbf{M}}\left[B^T(\boldsymbol{\xi})B(\boldsymbol{\xi})\right],\label{MARE6}\\
MV(\boldsymbol{\xi})&=&\frac{1}{N}\sum_{i=1}^N E_{\mathbf{M}}[Var_{\hat{\boldsymbol{\beta}}}(f(\boldsymbol{x}_i;\hat{\boldsymbol{\beta}}))]. 
\label{MARE7}
\end{eqnarray}
\end{lemma}
 In the following subsections, we will discuss the maximization of $MMPE(\psi, \boldsymbol{\xi})$ over $\mathcal{F}$ for multiple linear regression and nonlinear regression models separately.

\subsection{Multiple Linear Regression Model}

When $f(\mathbf{x}; \boldsymbol{\beta})=\mathbf{z}^T(\boldsymbol{x})\boldsymbol{\beta}$, where $\mathbf{z}(\boldsymbol{x})$ is a vector function of $\mathbf{x}$, the model (\ref{MARE1}) becomes the following multiple linear regression model
\begin{eqnarray}
Y=\mathbf{z}^T(\boldsymbol{x})\boldsymbol{\beta}+{\varepsilon}.
\label{MARE9}
\end{eqnarray}
Denote $\mathbf{Z}=(\mathbf{z}^T(\boldsymbol{x}_i))_{i=1}^N$
 and let
$\mathbf{R}(\boldsymbol{\xi},\mathbf{M})$ be
\begin{eqnarray}
\mathbf{R}(\boldsymbol{\xi},\mathbf{M})=\mathbf{Z}\left(\mathbf{Z}^T\mathbf{D}_{\boldsymbol{\xi}\mathbf{M}}\mathbf{Z} \right)^{-1} \mathbf{Z}^T,
\nonumber
\end{eqnarray}
with $\mathbf{D}_{\boldsymbol{\xi}\mathbf{M}}$ being the diagonal matrix of $\left(\sum_{j=1}^{n\xi_1}m_{1j},..., \sum_{j=1}^{n\xi_N}m_{Nj}\right).$ 
Notice that $\mathbf{D}_{\boldsymbol{\xi}\mathbf{M}}=\mathbf{D}_{\boldsymbol{\xi}} $ when $m_{ij}=1$ for all $j\leq n_i$ and $i=1,...,N.$ 

With the notations introduced above, Lemma \ref{decopos multi linear} below shows the explicit form of (\ref{MARE5}) for the multiple linear regression models.  

\begin{lemma}\label{decopos multi linear}
For the multiple linear regression model, the  optimality criterion MMPE is
\begin{eqnarray}
MMPE(\psi, \boldsymbol{\xi})
&=& \frac{1}{Nn} \Psi^T(\boldsymbol{\beta})\left\{E_{\mathbf{M}}\left[ \mathbf{D}_{\boldsymbol{\xi M}}\mathbf{R}^2(\boldsymbol{\xi}, \mathbf{M})\mathbf{D}_{\boldsymbol{\xi M}}\right]+\mathbf{I}\right\}\Psi(\boldsymbol{\beta}) \nonumber\\
&&+  \frac{1}{N}\sigma^2E_{ \mathbf{M}}\left[\boldsymbol{tr}(\mathbf{R}(\boldsymbol{\xi},\mathbf{M})) \right],
\label{MARE14}
\end{eqnarray}
where $\sigma^2$ is the variance of the response variable and $\boldsymbol{tr}(\mathbf{R}(\boldsymbol{\xi},\mathbf{M}))$ is the trace of the matrix  $\mathbf{R}(\boldsymbol{\xi},\mathbf{M}).$ \end{lemma}

Based on Lemma \ref{decopos multi linear}, we can maximize  $MMPE(\psi, \boldsymbol{\xi})
$  over the neighborhood $\mathcal{F}$ of $\Psi(\boldsymbol{\beta})$  as follows.
  
  \begin{theorem}\label{MARET1}
 For the multiple linear regression model, 
 the maximized $MMPE(\psi, \boldsymbol{\xi})$ over the neighborhood $\mathcal{F}$ of $\Psi(\boldsymbol{\beta})$ is 
\begin{eqnarray}
\max_{\mathbf{v\in }\mathbb{R}^{N-p}:||\mathbf{v}||\leq \eta }MMPE(\psi, \boldsymbol{\xi})&=&\frac{\eta^2}{Nn} E_{\mathbf{M}}\left[Ch_{max}\left( \mathbf{D}_{\boldsymbol{\xi M}}\mathbf{R}^2(\boldsymbol{\xi}, \mathbf{M})\mathbf{D}_{\boldsymbol{\xi M}}\right) \right] + \frac{\eta^2}{Nn}\nonumber\\
&&+\frac{\sigma^2}{N} E_{ \mathbf{M}}\left[\boldsymbol{tr}\left(\mathbf{R}(\boldsymbol{\xi},\mathbf{M})\right) \right].  
\label{MARE28}
\end{eqnarray}
Here $Ch_{max}(\mathbf{A})$ denotes the maximum eigenvalue of a matrix $\mathbf{A}.$
\end{theorem}
The difficulty of calculating the expectations directly in  (\ref{MARE28}) leads us to consider its  
 Taylor approximation,  and an approximation of the  loss function for the robust design is given in Theorem \ref{loss for multiple} below. 
 \begin{theorem}\label{loss for multiple} 
 For the multiple linear regression model, 
 (\ref{MARE28}) is approximately equal to 
 the loss function $\mathcal{L}_{\eta^2,\sigma^2}(\boldsymbol{\xi})$ which is defined  as follows
\begin{eqnarray}
\mathcal{L}_{\eta^2,\sigma^2}(\boldsymbol{\xi})&=&\frac{\eta^2}{Nn} Ch_{max}(\mathbf{D}_{\boldsymbol{\xi}}\mathbf{R}^2(\boldsymbol{\xi})\mathbf{D}_{\boldsymbol{\xi}})+\frac{\sigma^2}{N}\boldsymbol{tr}\left[\mathbf{R}(\boldsymbol{\xi})\right]+\frac{\eta^2}{Nn} \nonumber\\
&&- \frac{2\eta^2}{Nn} \mathbf{v}_1^T\mathbf{R}(\boldsymbol{\xi})\mathbf{D}_{\boldsymbol{\xi}}(\mathbf{I}-\mathbf{P}) \mathbf{v}_1 +\frac{2\eta^2}{Nn} \mathbf{v}_1^T \mathbf{R}(\boldsymbol{\xi})\mathbf{D}_{\boldsymbol{\xi}}(\mathbf{I}-\mathbf{P}) \mathbf{R}(\boldsymbol{\xi})\mathbf{D}_{\boldsymbol{\xi}}\mathbf{v}_1 \nonumber\\
&& + \frac{\sigma^2}{N}\boldsymbol{tr}\left[ (\mathbf{I}-\mathbf{P})\mathbf{D}_{\boldsymbol{\xi}} \mathbf{R}^2(\boldsymbol{\xi})\right].
\label{MARE42}
\end{eqnarray}
Here $\mathbf{R}(\boldsymbol{\xi})=\mathbf{Z} (\mathbf{Z}^T\mathbf{D}_{\boldsymbol{\xi}}\mathbf{Z})^{-1}\mathbf{Z}^T$,
 $\mathbf{P}$ is the diagonal matrix of the  vector $(p(m_i=1|\mathbf{x}_i,\gamma))_{i=1}^N$ with $p(m_i=1|\mathbf{x}_i,\gamma)$ being the response probability at $\mathbf{x}_i$, 
 and $\mathbf{v}_1$ is the normalized eigenvector of $Ch_{max}(\mathbf{D}_{\boldsymbol{\xi}}\mathbf{R}^2(\boldsymbol{\xi})\mathbf{D}_{\boldsymbol{\xi}})$.

\end{theorem}

\subsection{Nonlinear Regression Model}

In this section, we consider a nonlinear regression model (\ref{MARE1}) where $f(\mathbf{x}_i;\boldsymbol{\beta})$ is a smooth enough nonlinear function with bounded second derivatives. The procedure of looking for the approximated loss function and its maximum over $\mathcal{F}$ is succinctly described as follows.  To calculate the expectation with respect to the maximum likelihood estimate $\hat{\boldsymbol{\beta}}$ in the loss function,   we first derive the asymptotic distribution of  $f(\mathbf{x},\hat{\boldsymbol{\beta}})$  conditional on the missing indicators in Lemma \ref{asymaptotic distri}. Based on this result we can determine the asymptotic approximation of the loss function.
In Theorem \ref{MARET2} we maximize the approximated loss function over the neighborhood $\mathcal{F}$, and then the terms of the maximized loss will be Taylor expanded for the sake of calculating the expectations with respect to the missing indicator.   

\begin{lemma}\label{asymaptotic distri} Under the assumption that the missing indicators satisfy $$\sum_{i=1}^N\sum_{j=1}^{n_i}(1-m_{ij})=O_p(1),$$ for the nonlinear regression model (\ref{MARE1}),
$\sqrt{n}(f(\mathbf{x},\hat{\boldsymbol{\beta}})-f(\mathbf{x},{\boldsymbol{\beta}}))$ follows an asymptotic normal distribution with asymptotic mean 
\begin{eqnarray}
\mathbf{z}_i^T(\boldsymbol{\beta})(\mathbf{Z}^T(\boldsymbol{\beta})\mathbf{D}_{\boldsymbol{\xi M}}\mathbf{Z}(\boldsymbol{\beta}))^{-1}\mathbf{Z}^T(\boldsymbol{\beta}) \mathbf{D}_{\boldsymbol{\xi M}}  \boldsymbol{\Psi}(\boldsymbol{\beta})
\label{MARE32}
\end{eqnarray}
and asymptotic variance
\begin{eqnarray}
n\sigma^2 \mathbf{z}_i^T(\boldsymbol{\beta})(\mathbf{Z}^T(\boldsymbol{\beta})\mathbf{D}_{\boldsymbol{\xi M}}\mathbf{Z}(\boldsymbol{\beta}))^{-1}\mathbf{z}_i(\boldsymbol{\beta}).
\label{MARE33}
\end{eqnarray}

\end{lemma} 

According to the asymptotic distribution of $f(\mathbf{x},\hat{\boldsymbol{\beta}})$ shown in Lemma \ref{asymaptotic distri},   
the bias vector $B(\boldsymbol{\beta},\boldsymbol{\xi},\mathbf{M})$, asymptotically, becomes 
$$n^{-1/2}\mathbf{R}(\boldsymbol{\beta};\boldsymbol{\xi}, \mathbf{M})\mathbf{D}_{\boldsymbol{\xi M}} \boldsymbol{\Psi}(\boldsymbol{\beta}),$$
where
$\mathbf{R}(\boldsymbol{\beta};\boldsymbol{\xi}, \mathbf{M})=\mathbf{Z}(\boldsymbol{\beta}) (\mathbf{Z}^T(\boldsymbol{\beta})\mathbf{D}_{\boldsymbol{\xi M}}\mathbf{Z}(\boldsymbol{\beta}))^{-1}\mathbf{Z}^T(\boldsymbol{\beta})$.
Then 
\begin{eqnarray}
N\times MB(\boldsymbol{\xi})&=&E_{\mathbf{M}}[B^T(\boldsymbol{\beta},\boldsymbol{\xi},\mathbf{M})B(\boldsymbol{\beta},\boldsymbol{\xi},\mathbf{M}) ]\nonumber\\
&\approx&n^{-1}\boldsymbol{\Psi}^T(\boldsymbol{\beta})E_{\mathbf{M}}[\mathbf{D}_{\boldsymbol{\xi M}}\mathbf{R}^2(\boldsymbol{\beta}; \boldsymbol{\xi}, \mathbf{M})\mathbf{D}_{\boldsymbol{\xi M}}]\boldsymbol{\Psi}(\boldsymbol{\beta})
\label{MARE31}
\end{eqnarray}
and 
\begin{eqnarray}
 N\times MV(\boldsymbol{\xi})&=& E_{\mathbf{M}}\{\boldsymbol{tr}[Var_{\hat{\boldsymbol{\beta}}}(\mathbf{F}(\hat{\boldsymbol{\beta}}))]\}  \nonumber\\
 &\approx&\sigma^2 E_{ \mathbf{M}}  \left\{ \boldsymbol{tr}  \left[\mathbf{R}(\boldsymbol{\beta},\boldsymbol{\xi},\mathbf{M})\right] \right\}.
\nonumber
\end{eqnarray}
Thus the asymptotic $MMPE(\psi, \boldsymbol{\xi})$  for the nonlinear regression model becomes
\begin{eqnarray}
MMPE(\psi, \boldsymbol{\xi})
&=& \frac{1}{Nn} \Psi^T(\boldsymbol{\beta})\left\{E_{\mathbf{M}}\left[ \mathbf{D}_{\boldsymbol{\xi M}}\mathbf{R}^2(\boldsymbol{\beta};\boldsymbol{\xi}, \mathbf{M})\mathbf{D}_{\boldsymbol{\xi M}}\right]+\mathbf{I}\right\}\Psi(\boldsymbol{\beta}) \nonumber\\
&&+  \frac{1}{N}\sigma^2E_{ \mathbf{M}}\left[\boldsymbol{tr}\left(\mathbf{R}(\boldsymbol{\beta};\boldsymbol{\xi,M})\right) \right].
\label{MARE144}
\end{eqnarray}
The asymptotic loss function (\ref{MARE144}) for nonlinear models is similar to the loss function (\ref{MARE14}) for linear models except that (\ref{MARE144}) depends on the unknown regression parameter $\boldsymbol{\beta}$. In the following theorem, we maximize the asymptotic loss function (\ref{MARE144}) over $\mathcal{F}$.  The proof follows a  similar argument as for  Theorem \ref{MARET1}, and thus it is omitted.

\begin{theorem}\label{MARET2}
For the nonlinear regression model (\ref{MARE1}), the maximized $MMPE(\psi, \boldsymbol{\xi})$ over the neighborhood $\mathcal{F}$ of $\Psi(\boldsymbol{\beta})$ is 
\begin{eqnarray}
\max_{\mathbf{v\in }\mathbb{R}^{N-p}:||\mathbf{v}||\leq \eta }MMPE(\psi, \boldsymbol{\xi})
&=&\frac{\eta^2}{Nn} E_{\mathbf{M}}\left[ Ch_{max}\left( \mathbf{D}_{\boldsymbol{\xi M}}\mathbf{R}^2(\boldsymbol{\beta};\boldsymbol{\xi}, \mathbf{M})\mathbf{D}_{\boldsymbol{\xi M}}\right) \right] \nonumber\\
&&+ \frac{\eta^2}{Nn}+\frac{\sigma^2}{N} E_{ \mathbf{M}}\left[\boldsymbol{tr}\left(\mathbf{R}(\boldsymbol{\beta};\boldsymbol{\xi},\mathbf{M})\right) \right] .
\label{MARE299}
\end{eqnarray}

\end{theorem}

Applying Taylor expansion to (\ref{MARE299}), we will  end up with an approximated design criterion similar to  (\ref{MARE42}). 
The new design criterion for nonlinear regression model is denoted as $\mathcal{L}_{\eta^2,\sigma^2}(\boldsymbol{\xi};\boldsymbol{\beta})$,   which indicates that the model parameters $\boldsymbol{\beta}$ are included
 \begin{eqnarray}
\mathcal{L}_{\eta^2,\sigma^2}(\boldsymbol{\xi};\boldsymbol{\beta})&=& \frac{\eta^2}{Nn} Ch_{max}(\mathbf{D}_{\boldsymbol{\xi}}\mathbf{R}^2(\boldsymbol{\beta}; \boldsymbol{\xi })\mathbf{D}_{\boldsymbol{\xi}})+\frac{\sigma^2}{N}\boldsymbol{tr}\left[\mathbf{R}(\boldsymbol{\beta}; \boldsymbol{\xi })\right] \nonumber\\
&&+\frac{\eta^2}{Nn}- \frac{2\eta^2}{Nn} \mathbf{v}_1^T\mathbf{R}(\boldsymbol{\beta}; \boldsymbol{\xi })\mathbf{D}_{\boldsymbol{\xi}}(\mathbf{I}-\mathbf{P}) \mathbf{v}_1 \nonumber\\
&&+\frac{2\eta^2}{Nn} \mathbf{v}_1^T \mathbf{R}(\boldsymbol{\beta}; \boldsymbol{\xi })\mathbf{D}_{\boldsymbol{\xi}}(\mathbf{I}-\mathbf{P}) \mathbf{R}(\boldsymbol{\beta}; \boldsymbol{\xi })\mathbf{D}_{\boldsymbol{\xi}}\mathbf{v}_1\nonumber\\
&& + \frac{\sigma^2}{N}\boldsymbol{tr}\left[ (\mathbf{I}-\mathbf{P})\mathbf{D}_{\boldsymbol{\xi}} \mathbf{R}^2(\boldsymbol{\beta}; \boldsymbol{\xi })\right],
\label{MARE422}
\end{eqnarray}
where 
$\mathbf{P}$ is the diagonal matrix of the response probability vector $(p(m_i=1|\mathbf{x}_i,\gamma))_{i=1}^N$, 
and $\mathbf{v}_1$ is the normalized eigenvector of $Ch_{max}(\mathbf{D}_{\boldsymbol{\xi}}\mathbf{R}^2(\boldsymbol{\beta};\boldsymbol{\xi })\mathbf{D}_{\boldsymbol{\xi}})$.

In the next section, we will discuss how to obtain the minimax optimal designs by minimizing the maximized loss with respect to design $\boldsymbol{\xi}$.

\section{Robust optimal designs}\label{Sec numeric}

To obtain robust optimal designs, we minimize the design criterion (\ref{MARE42})  for multiple linear regression models, and (\ref{MARE422})  for nonlinear regression models. The minimization is completed by applying a genetic algorithm which has been developed using notions of evolutionary theory and widely used in literature. See, for example, \cite{MJWB2007}, \cite{WW2013}, \cite{ZW2015}. The  genetic algorithm applied here is a modification of that of \cite{KW2014}. We therefore only succinctly describe the general features of the algorithm as following. 
\begin{enumerate}
    \item An initial generation of a population of designs is first generated randomly  with size $n_g=40$. 
\item A ``fitness level'' for a design is defined in a way such that a design having a smaller value of loss is more fit.  For the current generation, the fitness levels are calculated for all the designs. The $N_{elite}=n_gP_{elite}$ elite (i.e., the most ``fit'') designs of the current generation pass through  to the next generation. Here $P_{elite}$ is a probability determined by the user.

\item Scaled fitness levels, which are proportional to the fitness values are used as selection probabilities to choose ``parent'' designs.  The  selected parent designs then produce ``children'' via stochastic processes of ``crossover'' and ``mutation''. 
The process is repeated, until the current generation of $n_g$ designs has been replaced by a new generation which consists of $n_g-N_{elite}$ child designs and $N_{elite}$ the most fit designs.    
\item The inclusion of the elite members guarantees the loss non-increasing in each generation. The algorithm terminates when the minimum loss has not changed in 200 consecutive generations. 

\end{enumerate}

We run the genetic algorithm described above in the following Example 1 to Example 4. In each example, minimax optimal designs will be obtained for various sets of parameter values. For example, the value of $\eta^2$ will be changed so we can study the effect of the size of neighborhoods $\mathcal{F}$ on the robust designs.  As a comparison, ``classical'' optimal designs will be obtained by letting $\eta^2=0$ in all the cases. These designs are referred as ``classical''  because when $\eta^2=0$,  the proposed models are assumed to be the true models, which is a common assumption in the classical optimal design theory.  

\subsection{Multiple Linear Regression}

 \noindent\textbf{Example 1: A polynomial model}
Consider a design space, which includes 100 equally spaced  design points  in  $[0,1]$.  The regression function proposed by the researcher is
\begin{eqnarray}
f({x}; \boldsymbol{\beta})=\beta_0+\beta_1x+\beta_2x^2+\beta_3x^3.
\label{MARE29}
\end{eqnarray}
As an illustration, we assume that the missing indicator has the following probability 
$$p(m_i=1|x_i, {\gamma}_0=2, \gamma_1=0.5)=\frac{exp(0.5x_i+2)}{1+exp(0.5x_i+2)},~i=1,2,...,50.$$

\begin{figure}[h!]
    \centering
    \includegraphics[width=11cm]{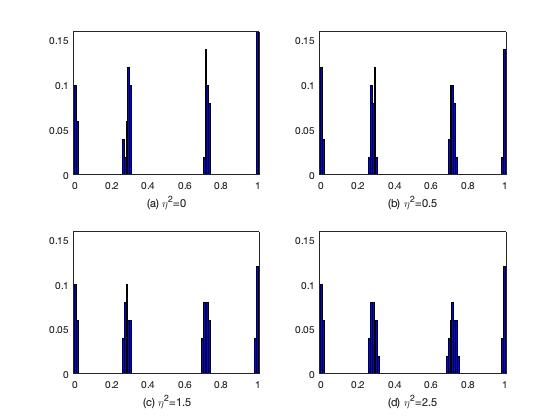}    \caption{Optimal designs for Example 1 with $\sigma^2=0.01$ and varying $\eta^2$. Relative frequencies of the design points used in the design vs the design points.}
\label{MARF1}
\end{figure}

As shown in Figure \ref{MARF1} (a),  an optimal design 
is obtained with $\eta^2=0$, which indicates that this design does not attempt robustness within $\mathcal{F}$. To illustrate the effect of $\eta^2$, i.e., the uncertainty of the correct model, we let the value $\eta^2$ vary from 0.5 to 2.5. When the value of $\eta^2$ increases, the  neighborhood $\mathcal{F}$ is getting larger and the uncertainty of the correct model also increases.  Figures \ref{MARF1} (b)-(c) show the minimax optimal designs obtained for three different values of $\eta^2$ ($=0.5, 1.5, 2.5$, respectively). Comparing the four designs in Figure \ref{MARF1}, we find that the minimax designs tend to be more clustered than the classical optimal design, and  as $\eta^2$ increases they also become more dispersed. 

The variance $\sigma^2$ of the response variable is set to 0.01 in  Figure \ref{MARF1}. We also obtained the nonrobust and minimax  designs for different values of $\sigma^2$. The designs have the properties that show in Figure \ref{MARF1}, i.e., as the uncertainty about the correct model increases the optimal design tends to be less clustered as a protection against possible model mis-specification. The plots of the designs are omitted to save space. However, for a fixed $\eta^2$ the value of maximum loss for the robust design increases with respect to $\sigma^2$ as shown in Table \ref{MART1}. From this table, we can also observe a natural result that the value of the maximum loss increases as $\eta^2$ increases. 

\begin{table}[h!]
\centering
\begin{tabular}{cccccccccccccccc}
\hline
 &  & \multicolumn{11}{c}{ $\eta^2$} \\ \cline{3-13}
    $\sigma^2$      &&0.3 &&  0.5       && 1 &&     1.5        && 2 && 2.5   \\ \hline
$0.01$ && 0.7304   && 0.7670    &&  0.8726  &&  0.9700   &&  1.0672   &&   1.1703     \\ 
$0.05$&&  3.4071  &&  3.4484  && 3.5487  && 3.6498    &&  3.7306  && 3.8608   \\
0.1 && 6.7437  && 6.7558 && 6.8664 && 6.9677 && 7.1026 && 7.2155 \\ \hline
\end{tabular}
  \caption{ Components of loss $\times 10^3$ for Example 1 with various $\sigma^2$ and $\eta^2$}
    \label{MART1}
\end{table}

\noindent\textbf{Example 2: A linear regression model with two predictors}
Consider a design space which includes 100 equally spaced  design points  in a square $[0,1]\times [0,1]$. We consider the following model 
\begin{eqnarray}
Y=\beta_0+\beta_1x_1+\beta_2x_2+\beta_3x_1x_2+\beta_4x_1^2+\beta_5x_2^2+\varepsilon.
\label{MARE43}
\end{eqnarray}
The missing indicator is assumed to have the following probability 
$$p(m_i=1|{x}_{i1},x_{i2}, {\gamma}_0=2, \gamma_1=0.5, \gamma_2=0.5)=\frac{\exp(0.5x_{i1}+0.5x_{i2}+2)}{1+\exp(0.5x_{i1}+0.5x_{i2}+2)}.$$
 
\begin{figure}[h!]
    \centering
    \includegraphics[width=11cm]{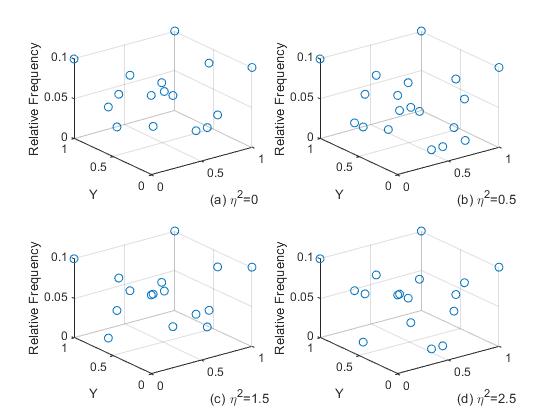}
    \caption{Classical optimal design and minimax optimal designs for the multiple model (\ref{MARE43}) with $\sigma^2=0.01$. } 
    \label{MARF3}
\end{figure}

In Figure \ref{MARF3}, we obtained the minimax optimal designs with $\sigma^2=0.01$ and $\eta^2=0, 0.5, 1.5, 2.5$, respectively. The $(X,Y)-$coordinates of a blue circle in the plots is the location of the selected design point and the $Z$ coordinate is the relative frequency of the design point in the design.  To investigate the effect of the variance of the response variable, we let $\sigma^2$ vary from 0.01, 0.05 to 0.1. Similar properties as in Example 1 are observed,  and  from the results in Table \ref{MART2} the maximum loss increases as the values of $\sigma^2$ and $\eta^2$ are getting larger.

\begin{table}[h!]
\centering
\begin{tabular}{cccccccccccccccc}
\hline
 &  & \multicolumn{11}{c}{ $\eta^2$} \\ \cline{3-13}
    $\sigma^2$      &&0.3 &&  0.5       && 1 &&     1.5        && 2 && 2.5   \\ \hline
$0.01$ && 0.9613  &&  1.0037   && 1.1099 &&  1.2018  && 1.3013  &&  1.4017      \\ 
$0.05$&& 4.5638 && 4.6013  && 4.7902  &&  4.8015  && 4.8976 && 5.0074  \\
0.1 && 9.0880  && 9.0916 && 9.2355 && 9.3498  &&  9.4014 &&  9.5273  \\ \hline
\end{tabular}
  \caption{Components of loss $\times 10^3$ for Example 2  with various $\sigma^2$ and $\eta^2$}
    \label{MART2}
\end{table}

\subsection{Nonlinear Regression}

The design criterion for a nonlinear regression depends on unknown regression parameters $\boldsymbol{\beta}$. Our method to handle the dependency on the unknown parameters is through the Bayesian-optimal mechanism. We first average the design criterion with respect to a ``prior'' distribution, and then minimizing the averaged design criterion over all the possible designs.

 \bigskip
\noindent\textbf{Example 3: A nonlinear regression model and Bayesian optimality} 
We consider the model for long-term recovery after discharge from the hospitals in \cite{KNN2003}, a  two-parameter nonlinear regression model with $f(x|\boldsymbol{\beta})=\beta_0\exp\{\beta_1x\}$ where $x$ is  the number of days of hospitalization. Let $T_1$ and $T_2$ be two new variables such that
$$\beta_0=57(T_1+0.5), \hspace{10pt} \beta_1=-\frac{T_2+.5}{25}.$$
By performing a preliminary linear regression  with response, $\log(Y)$, and predictor $X$, estimates of $\hat{\boldsymbol{\beta}}=(56.7, -.03797)$ were obtained. So, we let the values of $T_1$ and $T_2$ vary between 0 and 1  and 
$\beta_0\in[28.5,85.5], \hspace{10pt} \beta_1\in [-.06,-.02]$.
Moreover, assume that the  prior distributions of $T_1$ and $T_2$ have densities $p(t_1;\alpha_1,\alpha_2)$ and $ p(t_2;\alpha_1,\alpha_2)$, respectively. Then the design criterion (\ref{MARE28})  is replaced by
\begin{eqnarray}
 \int_0^1\int_0^1 \mathcal{L}_{\eta^2,\sigma^2}(\boldsymbol{\xi};t_1,t_2) p(t_1;\alpha_1,\alpha_2)p(t_2;\alpha_1,\alpha_2)dt_1dt_2.
\label{MARE44}
\end{eqnarray}

\begin{figure}[h!]
     \centering
     \includegraphics[width=11cm]{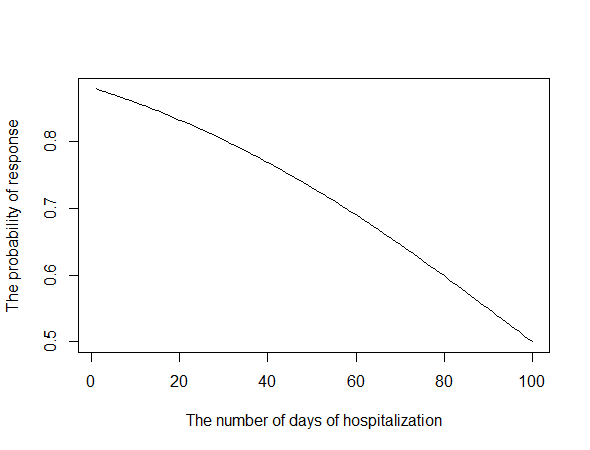}
     \caption{The plot of the probability of response against days $x\in [1,100]$}
     \label{MARF7}
 \end{figure}

The design space includes $\{1, 2, ..., 100\}$, which are the numbers of days of hospitalization. Since missing data is more and more likely to occur as days go by, the probability of response should be decreasing with respect to $x$. In this example, we assume that the probability of response is 
$$p(m_i=1|x_i, \gamma_0=2, \gamma_1=-0.02)=\frac{\exp(-0.02x_i+2)}{1+\exp(-0.02x_i+2)}.$$
 Figure \ref{MARF7} shows that this probability is a decreasing function of $x$. When $x=1$, the probability of response is  0.88; while for $x=100$, the probability decreases to 0.5. 
 
In this example, besides the discussion about the change of the minimax designs with respect to varied sizes of the neighborhood $\mathcal{F}$,  we will also investigate the effect of the prior distribution assumption of the regression parameters. Since  both $T_1$ and $T_2$ are in $[0,1]$, reasonable prior distributions for these two random variables are the uniform distribution on $[0,1]$ and Beta distributions. We will study the following three scenarios: (a) both $T_1$ and $T_2$ have uniform distributions on $[0,1]$; (b) both $T_1$ and $T_2$ have symmetric Beta distributions Beta$[5,5]$; (c) both $T_1$ and $T_2$ have right-skewed Beta distributions Beta$[2,4]$. In these three cases, the variance of the response is set to $\sigma^2=0.01$, and the size of $\mathcal{F}$ varies as $\eta^2=0, .5, 1.5, 2.5$, respectively.

\begin{figure}[h!]
    \centering
    \includegraphics[width=11cm]{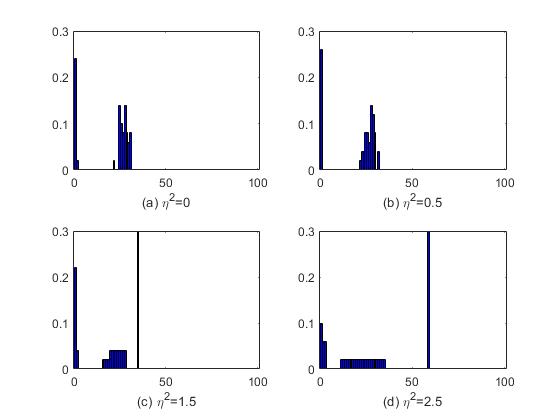}
    \caption{For the nonlinear model $f(x|\boldsymbol{\beta})$ of Example 3 with $\sigma^2=0.01$ and Uniform$[0,1]$ prior} 
    \label{MARF5}
\end{figure}

\begin{figure}[h!]
    \centering
    \includegraphics[width=11cm]{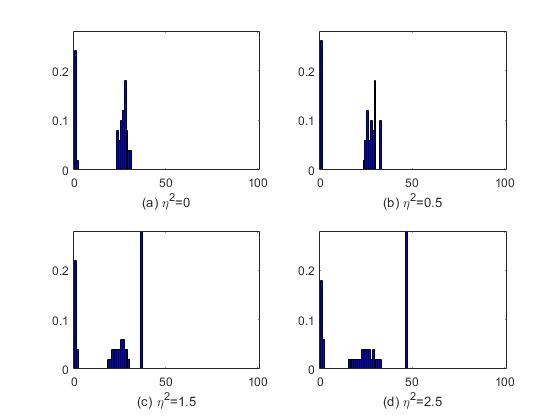}
    \caption{For the nonlinear model $f(x|\boldsymbol{\beta})$ of Example 3 with $\sigma^2=0.01$ and Beta$[5,5]$ prior.} 
    \label{MARF6}
\end{figure}

\begin{figure}[h!]
    \centering
    \includegraphics[width=11cm]{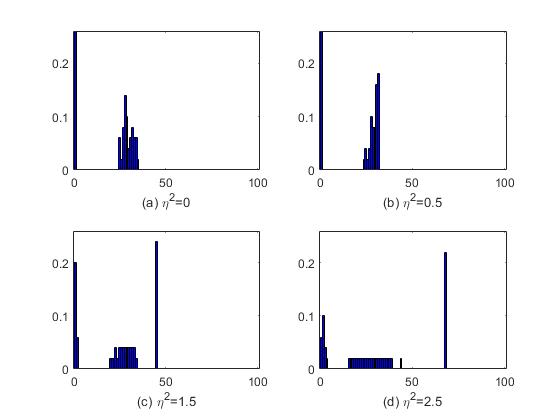}
    \caption{For the nonlinear model $f(x|\boldsymbol{\beta})$ of Example 3 with $\sigma^2=0.01$ and Beta$[2,4]$ prior.} 
    \label{MARF8}
\end{figure}

The optimal designs shown in Figures \ref{MARF5}-\ref{MARF8} are obtained with the Unif(0,1) prior, Beta$[5,5]$ prior, and Beta$[2,4]$ prior, respectively. It is no surprise to see that for fixed prior distributions the optimal designs become more dispersed as the value of $\eta^2$ increases from 0 to 2.5.  It is noticeable that the shape of the prior distribution doesn't have very significant effect on the optimal designs because the designs with different priors do not have very critical differences. However, on the maximum loss of the minimax optimal designs prior distributions have effects that cannot be ignored. Although both of the Unif(0,1) and Beta[5,5] are the symmetric prior distributions, the maximum losses with Unif(0,1) are always larger than those with Beta[5,5]. For Beta[2,4], a right-skewed prior distribution, its maximum losses are very close to those of Beta[5,5].  According to the above analysis, we can conclude that the prior distributions have very limited effect on the optimal designs, and the value of the maximum loss does not depend on the parameters of the prior distribution. But the maximum loss of the minimax designs can be very different for distinct prior distributions.

\begin{table}[h!]
\centering
\begin{tabular}{cccccccccccccccc}
\hline
 &  & \multicolumn{11}{c}{ $\eta^2$} \\ \cline{3-13}
          &&0.3 &&  0.5       && 1 &&     1.5        && 2 && 2.5   \\ \hline
Unif(0,1) && 0.4269   &&  0.4875    && 0.6388  &&  0.7832  && 0.9306  &&  1.0722      \\ 
Beta(2,4)&&  0.2935 && 0.3275 &&  0.4263 &&  0.5242  && 0.6176  && 0.7080   \\ 
Beta(5,5) && 0.3001    &&  0.3406 && 0.4451   && 0.5496  && 0.6536  &&  0.7516 \\ \hline   
\end{tabular}
  \caption{Components of loss $\times 10^3$ for Example 3  with various  $\eta^2$ and prior distributions}
    \label{MART3}
\end{table}

\section{Summary and concluding remarks}

This paper proposes obtaining methods for minimax optimal designs for use with possible missing data when the model's functional form is in doubt. We first proposed a neighbourhood of the parametric model thought to be a reasonable approximation to the true response. Within this neighborhood, we then maximized the design criterion $MMPE$, the \textbf{M}ean of averaged \textbf{M}eans of squared \textbf{P}rediction \textbf{E}rrors, with respect to the missing indicator. The loss as a function of the design vector,  the minimax optimal design could be found by minimizing the maximized loss function. But the loss may also depend on the model parameters. In this case, we integrated out the unknown parameters in the loss by applying the Bayesian method. The result of this process, which only depends on the designs, was then minimized by implementing the generic algorithm using Matlab to look for the optimal designs. We studied three examples whose proposed models were polynomial, multiple linear and nonlinear regression models, respectively. The ``classical'' optimal designs with $\eta^2=0$ and minimax optimal designs with nonzero $\eta^2$ were obtained and compared. We also studied the effects of the variance of the response and the size of the neighborhood $\mathcal{F}$ by varying the values of $\sigma^2$ and $\eta^2$. We discovered that the optimal designs had fewer support points when the neighbourhood was small, thus more replicates. But when the size of the neighbourhood increased, the design points were becoming more and more scattered. The variance of the response had a noticeable effect on the maximum loss of the minimax optimal designs, which increased as the value of $\sigma^2$ was getting larger. For the nonlinear regression, we also discussed the effect induced by the prior distribution. We consider three different prior distributions: two symmetric distributions (Unif(0,1) and Beta[5, 5]), and one asymmetric distribution (Beta[2, 4]). We could observe that the designs were not dramatically different for distinct prior distributions, but maximum losses' values become very different if the prior distribution changed.



\section*{Acknowledgements}
Hu's research is supported by the Natural Sciences and Engineering Research Council of Canada.
\par


\bibhang=1.7pc
\bibsep=2pt
\fontsize{9}{14pt plus.8pt minus .6pt}\selectfont
\renewcommand\bibname{\large \bf References}
\expandafter\ifx\csname
natexlab\endcsname\relax\def\natexlab#1{#1}\fi
\expandafter\ifx\csname url\endcsname\relax
  \def\url#1{\texttt{#1}}\fi
\expandafter\ifx\csname urlprefix\endcsname\relax\def\urlprefix{URL}\fi








\newpage

\section*{Appendix}


\subsection*{Appendix A. Proof of Theorem \ref{Throem decompse of loss} }

Multiplying by $N$ on both sides of (\ref{MARE8}), we have
\begin{eqnarray}
& &N\times MMPE(\psi, \boldsymbol{\xi})
\nonumber\\
&=& \sum_{i=1}^N E_{\hat{\boldsymbol{\beta}}, \mathbf{M}}\left[\left( f(\boldsymbol{x}_i;\hat{\boldsymbol{\beta}})-f(\boldsymbol{x}_i; \boldsymbol{\beta})-\frac{\psi(\boldsymbol{x}_i; \boldsymbol{\beta})}{\sqrt{n}}\right)^2\right]
\nonumber\\
&=& \sum_{i=1}^N E_{\hat{\boldsymbol{\beta}}, \mathbf{M}}\left[\left(f(\boldsymbol{x}_i; \hat{\boldsymbol{\beta}})-E_{\hat{\boldsymbol{\beta}}}(f(\boldsymbol{x}_i; \hat{\boldsymbol{\beta}}))+E_{\hat{\boldsymbol{\beta}}}(f(\boldsymbol{x}_i;\hat{\boldsymbol{\beta}}))-f(\boldsymbol{x}_i; \boldsymbol{\beta})-\frac{\psi(\boldsymbol{x}_i; \boldsymbol{\beta})}{\sqrt{n}}\right)^2\right]\nonumber\\
&=& E_{\mathbf{M}}\left[ \sum_{i=1}^N E_{\hat{\boldsymbol{\beta}}}[f(\boldsymbol{x}_i;\hat{\boldsymbol{\beta}})-E(f(\boldsymbol{x}_i;\hat{\boldsymbol{\beta}}))]^2+\sum_{i=1}^N [E_{\hat{\boldsymbol{\beta}}}(f(\boldsymbol{x}_i;\hat{\boldsymbol{\beta}}))-f(\boldsymbol{x}_i; \boldsymbol{\beta})]^2 \right.
\nonumber\\
& &\left.+\frac{\|{\Psi}(\boldsymbol{\beta})\|^2}{n} -2\sum_{i=1}^N[E_{\hat{\boldsymbol{\beta}}}(f(\boldsymbol{x}_i;\hat{\boldsymbol{\beta}}))-f(\boldsymbol{x}_i; \boldsymbol{\beta})] \frac{\psi(\mathbf{x}_i; \boldsymbol{\beta})}{\sqrt{n}}\right] \nonumber\\
&=&N\times MV(\boldsymbol{\xi})+N\times MB(\boldsymbol{\xi}) - 2E_{\mathbf{M}}({B}^T(\boldsymbol{\xi})\frac{{\Psi}(\boldsymbol{\beta}))}{\sqrt{n}}
+\frac{\|{\Psi}(\boldsymbol{\beta})\|^2}{n},
\label{MARE35}
\end{eqnarray}
where 
\begin{eqnarray*}
N\times MV(\boldsymbol{\xi})&=&\sum_{i=1}^N E_{\hat{\boldsymbol{\beta}},\mathbf{M}}[f(\boldsymbol{x}_i;\hat{\boldsymbol{\beta}})-E_{\hat{\boldsymbol{\beta}}}(f(\boldsymbol{x}_i;\hat{\boldsymbol{\beta}}))]^2\\
&=&\sum_{i=1}^N E_{\mathbf{M}}[\hbox{Var}_{\hat{\boldsymbol{\beta}}}f(\boldsymbol{x}_i;\hat{\boldsymbol{\beta}})]
\end{eqnarray*}
and 
$$ N\times MB(\boldsymbol{\xi})=E_{\mathbf{M}}\left[\sum_{i=1}^N[ E_{\hat{\boldsymbol{\beta}}}(f(\boldsymbol{x}_i;\hat{\boldsymbol{\beta}}))-f(\boldsymbol{x}_i; \boldsymbol{\beta})]^2 \right]=E_{\mathbf{M}}\left[B^T(\boldsymbol{\xi}) B(\boldsymbol{\xi})\right]. $$ 


\subsubsection*{Appendix B. Proof of Lemma \ref{decopos multi linear}}
The maximum likelihood estimate $\hat{\boldsymbol{\beta}}$ of the regression parameters of a multiple linear regression model (\ref{MARE9}) is  
\begin{eqnarray}
&&\hat{\boldsymbol{\beta}}=\left(\sum_{i=1}^N\sum_{j=1}^{n_i}m_{ij}\mathbf{z}_i\mathbf{z}_i^T\right)^{-1}\sum_{i=1}^N\sum_{j=1}^{n_i}m_{ij}\mathbf{z}_iy_{ij}\nonumber\\
&=& \left(\mathbf{Z}^T\mathbf{D}_{\boldsymbol{\xi}\mathbf{M}}\mathbf{Z} \right)^{-1} \mathbf{Z}^T
\left(\begin{array}{c}
     \sum_{j=1}^{n\xi_1} m_{1j}y_{1j}  \\
     ...\\
     \sum_{j=1}^{n\xi_N} m_{Nj}y_{Nj} 
\end{array} \right).
\nonumber
\end{eqnarray}
We then have 
\begin{eqnarray}
E[\hat{\boldsymbol{\beta}}]
&=&\left(\mathbf{Z}^T\mathbf{D}_{\boldsymbol{\xi}\mathbf{M}}\mathbf{Z} \right)^{-1} \mathbf{Z}^T
\left(\begin{array}{c}
     \sum_{j=1}^{n\xi_1} m_{1j}E[y_{1j}]  \\
     ...\\
     \sum_{j=1}^{n\xi_N} m_{Nj}E[y_{Nj}]
\end{array} \right)
\nonumber\\
&=& \left(\mathbf{Z}^T\mathbf{D}_{\boldsymbol{\xi}\mathbf{M}}\mathbf{Z} \right)^{-1} \mathbf{Z}^T
\left(\begin{array}{c}
     \sum_{j=1}^{n\xi_1} m_{1j}(\mathbf{z}_1^T\boldsymbol{\beta}+\frac{\psi(\mathbf{x}_1;\boldsymbol{\beta})}{\sqrt{n}})  \\
     ...\\
     \sum_{j=1}^{n\xi_N} m_{Nj}(\mathbf{z}_N^T\boldsymbol{\beta}+\frac{\psi(\mathbf{x}_N;\boldsymbol{\beta})}{\sqrt{n}}) 
\end{array} \right) \nonumber\\
&=& \left(\mathbf{Z}^T\mathbf{D}_{\boldsymbol{\xi}\mathbf{M}}\mathbf{Z} \right)^{-1} \mathbf{Z}^T\mathbf{D}_{\boldsymbol{\xi}\mathbf{M}}\left(\mathbf{Z}\boldsymbol{\beta}+\frac{{\Psi}(\boldsymbol{\beta})}{\sqrt{n}} \right),
\nonumber
\end{eqnarray}
and 
\begin{eqnarray}
Var[\hat{\boldsymbol{\beta}}]&=&\left(\mathbf{Z}^T\mathbf{D}_{\boldsymbol{\xi}\mathbf{M}}\mathbf{Z} \right)^{-1} \mathbf{Z}^T \hbox{diag}\left\{\sum_{j=1}^{n\xi_i} m_{ij}Var[y_{ij}]\right\}_{i=1}^N\mathbf{Z}\left(\mathbf{Z}^T\mathbf{D}_{\boldsymbol{\xi}\mathbf{M}}\mathbf{Z} \right)^{-1} 
\nonumber\\
&=& \sigma^2\left(\mathbf{Z}^T\mathbf{D}_{\boldsymbol{\xi}\mathbf{M}}\mathbf{Z} \right)^{-1}.
\nonumber
\end{eqnarray}

For a multiple linear regression model, the  bias vector $B(\boldsymbol{\xi}, \mathbf{M})$ becomes
\begin{eqnarray}
B(\boldsymbol{\xi},\mathbf{M})&=&
\mathbf{Z}E[\hat{\boldsymbol{\beta}}]-\mathbf{Z}{\boldsymbol{\beta}}
\nonumber\\
&=& \mathbf{Z}\left(\mathbf{Z}^T\mathbf{D}_{\boldsymbol{\xi}\mathbf{M}}\mathbf{Z}\right)^{-1} \mathbf{Z}^T\mathbf{D}_{\boldsymbol{\xi}\mathbf{M}}\left(\mathbf{Z}\boldsymbol{\beta}+\frac{{\Psi}(\boldsymbol{\beta})}{\sqrt{n}}\right)-\mathbf{Z}{\boldsymbol{\beta}}
\nonumber\\
&=& \mathbf{Z}\left(\mathbf{Z}^T\mathbf{D}_{\boldsymbol{\xi}\mathbf{M}}\mathbf{Z}\right)^{-1} \mathbf{Z}^T\mathbf{D}_{\boldsymbol{\xi M}}\frac{{\Psi}(\boldsymbol{\beta})}{\sqrt{n}}\nonumber\\
&=&   \mathbf{R}(\boldsymbol{\xi}, \mathbf{M}) \mathbf{D}_{\boldsymbol{\xi} \mathbf{M}}\frac{{\Psi}(\boldsymbol{\beta})}{\sqrt{n}},
\label{MARE12}
\end{eqnarray}
where $\mathbf{R}(\boldsymbol{\xi}, \mathbf{M})=\mathbf{Z}\left(\mathbf{Z}^T\mathbf{D}_{\boldsymbol{\xi}\mathbf{M}}\mathbf{Z}\right)^{-1} \mathbf{Z}^T$. We then have
$$B^T(\boldsymbol{\xi},\mathbf{M}) \Psi(\boldsymbol{\beta})=n^{-1/2}\Psi^T(\boldsymbol{\beta}) \mathbf{D}_{\boldsymbol{\xi}\mathbf{M}} \mathbf{Z}\left(\mathbf{Z}^T\mathbf{D}_{\boldsymbol{\xi}\mathbf{M}}\mathbf{Z}\right)^{-1} \mathbf{Z}^T\Psi(\boldsymbol{\beta})=0$$
according to the orthogonal condition. That is, the last term of (\ref{MARE35}) is 0. 
Moreover,
\begin{eqnarray}
&& N\times MB(\boldsymbol{\xi})\nonumber\\
&=&E_{\mathbf{M}}[B^T(\boldsymbol{\xi}, \mathbf{M})B(\boldsymbol{\xi}, \mathbf{M})]\nonumber\\
&=&n^{-1}\Psi^T(\boldsymbol{\beta})E_{\mathbf{M}}\left[\mathbf{D}_{\boldsymbol{\xi}\mathbf{M}} \mathbf{R}^2(\boldsymbol{\xi},\mathbf{M})\mathbf{D}_{\boldsymbol{\xi}\mathbf{M}}\right]\Psi(\boldsymbol{\beta}).
\label{MARE13}
\end{eqnarray}

Next, we find $MV(\boldsymbol{\xi})$:
\begin{eqnarray}
N\times MV(\boldsymbol{\xi})&=&
\sum_{i=1}^N E_{\hat{\boldsymbol{\beta}}, \mathbf{M}}[\mathbf{z}^T(\boldsymbol{x}_i)\hat{\boldsymbol{\beta}}-E(\mathbf{z}^T(\boldsymbol{x}_i)\hat{\boldsymbol{\beta}})]^2 \nonumber\\
&=& E_{\hat{\boldsymbol{\beta}}, \mathbf{M}}\left[\left(\mathbf{Z}\hat{\boldsymbol{\beta}} - \mathbf{Z}E(\hat{\boldsymbol{\beta}}) \right)^T\left(\mathbf{Z}\hat{\boldsymbol{\beta}} - \mathbf{Z}E(\hat{\boldsymbol{\beta}}) \right) \right]
\nonumber\\
&=& E_{\hat{\boldsymbol{\beta}}, \mathbf{M}}\left[\left(\hat{\boldsymbol{\beta}} - E(\hat{\boldsymbol{\beta}}) \right)^T\mathbf{Z}^T\mathbf{Z}\left(\hat{\boldsymbol{\beta}} - E(\hat{\boldsymbol{\beta}}) \right) \right]
\nonumber\\
&=& \boldsymbol{tr} \left\{ \mathbf{Z}^T\mathbf{Z} E_{\hat{\boldsymbol{\beta}}, \mathbf{M}}\left[\left(\hat{\boldsymbol{\beta}} - E(\hat{\boldsymbol{\beta}}) \right)\left(\hat{\boldsymbol{\beta}} - E(\hat{\boldsymbol{\beta}}) \right)^T \right] \right\}
\nonumber\\
&=& \boldsymbol{tr} \left\{ \mathbf{Z}^T\mathbf{Z} E_{ \mathbf{M}}\left[Var[\hat{\boldsymbol{\beta}}]\right] \right\}
\nonumber\\
&=& \sigma^2 E_{ \mathbf{M}} \left[ \boldsymbol{tr}  \left\{ \mathbf{R}(\boldsymbol{\xi},\mathbf{M})\right\}\right].
\nonumber
\end{eqnarray}

\subsection*{Appendix C. Proof of Theorem \ref{MARET1}}

Notice that the orthogonality requirement ${\Psi }^T(\boldsymbol{\beta})\mathbf{Z}=0$ is
equivalent to  $\Psi(\boldsymbol{\beta})$ lies in $R^{\bot}(\mathbf{Z})$.
Let $\mathbf{K}$ be an ${N\times(N-p)}$ matrix whose columns form an orthonormal basis for this
orthogonal complement, i.e., $\mathbf{K}^T\mathbf{K}=\mathbf{I}_{N-p}$. Then $\Psi(\boldsymbol{\beta})=\mathbf{Kv}$  for some $\mathbf{v}\in \mathbb{R}^{N-p}$ with $\|\mathbf{v}\|=\|\Psi(\boldsymbol{\beta}) \|$. Thus, maximizing  $MMPE(\psi, \boldsymbol{\xi})$ over $\Psi(\boldsymbol{\beta})$ is 
 equivalent to solving the following problem
\begin{eqnarray}
&&\max_{\mathbf{v\in }\mathbb{R}^{N-p}:||\mathbf{v}||\leq \eta }MMPE(\psi, \boldsymbol{\xi})\nonumber\\
&=&\max_{
\mathbf{v\in }\mathbb{R}^{N-p}:||\mathbf{v}||\leq \eta }\left\{\frac{1}{Nn} \mathbf{v}^T\mathbf{K}^T\left\{E\left[ \mathbf{D}_{\boldsymbol{\xi M}}\mathbf{R}^2(\boldsymbol{\xi}, \mathbf{M})\mathbf{D}_{\boldsymbol{\xi M}}\right]+\mathbf{I}\right\}\mathbf{Kv}\right\} \nonumber\\
&& +  \frac{1}{N}\sigma^2 E_{ \mathbf{M}}  \left\{ \boldsymbol{tr}  \left[\mathbf{R}(\boldsymbol{\xi},\mathbf{M})\right] \right\}.  \nonumber
\end{eqnarray}

We can decompose  $\mathbf{D}_{\boldsymbol{\xi M}}\mathbf{R}^2(\boldsymbol{\xi}, \mathbf{M})\mathbf{D}_{\boldsymbol{\xi M}}$ as $\mathbf{Q}\Lambda\mathbf{Q}^T$ where $\mathbf{Q}$ is  an orthogonal matrix of size $N\times N$ and $\Lambda$ is an $N\times N$ diagonal matrix with eigenvalues as its diagonal elements. Because $\|\mathbf{v}^T\mathbf{K}^T\mathbf{Q}\mathbf{Q}^T \mathbf{Kv} \|=\|\mathbf{v}\|$ we  have

\begin{eqnarray}
&&\max_{
\mathbf{v\in }\mathbb{R}^{N-p}:||\mathbf{v}||\leq \eta }\left\{\frac{1}{Nn}E\left[  \mathbf{v}^T\mathbf{K}^T\left\{\mathbf{D}_{\boldsymbol{\xi M}}\mathbf{R}^2(\boldsymbol{\xi}, \mathbf{M})\mathbf{D}_{\boldsymbol{\xi M}}\right\}\mathbf{Kv}\right]\right\}\nonumber\\
&\leq& \frac{1}{Nn} E_{\mathbf{M}}\left[ \max_{
\mathbf{v\in }\mathbb{R}^{N-p}:||\mathbf{v}||\leq \eta }\left\{\mathbf{v}^T\mathbf{K}^T\mathbf{Q}\Lambda\mathbf{Q}^T \mathbf{Kv}\right\}\right]\nonumber\\
&\leq& \frac{1}{Nn}  E\left[\max diag(\Lambda) \max_{
\mathbf{v\in }\mathbb{R}^{N-p}:||\mathbf{v}||\leq \eta }\left\{ \mathbf{v}^T\mathbf{K}^T\mathbf{Q}\mathbf{Q}^T \mathbf{Kv}\right\}\right]\nonumber\\
&=& \frac{1}{Nn} E\left[ \max diag(\Lambda) \max_{
\mathbf{v\in }\mathbb{R}^{N-p}:||\mathbf{v}||\leq \eta }\|\mathbf{v}\|^2\right]\nonumber\\
&= & \frac{\eta^2}{Nn} E\left[ Ch_{max}\left[ \mathbf{D}_{\boldsymbol{\xi M}}\mathbf{R}^2(\boldsymbol{\xi}, \mathbf{M})\mathbf{D}_{\boldsymbol{\xi M}}\right] \right],
\nonumber
\end{eqnarray}
where $Ch_{max}$ of a matrix denotes the largest eigenvalue of the matrix.

Therefore,
\begin{eqnarray}
&&\max_{\mathbf{v\in }\mathbb{R}^{N-p}:||\mathbf{v}||\leq \eta }MMPE(\psi, \boldsymbol{\xi})\nonumber\\
&=&\frac{\eta^2}{Nn} E_{ \mathbf{M}}\left[ Ch_{max}\left( \mathbf{D}_{\boldsymbol{\xi M}}\mathbf{R}^2(\boldsymbol{\xi}, \mathbf{M})\mathbf{D}_{\boldsymbol{\xi M}}\right) \right] + \frac{\eta^2}{Nn}\nonumber\\
&&+\frac{\sigma^2}{N}  E_{ \mathbf{M}}  \left\{ \boldsymbol{tr}  \left[\mathbf{R}(\boldsymbol{\xi},\mathbf{M})\right] \right\}.  \nonumber
\end{eqnarray}

\subsection*{Appendix D. Proof of Theorem \ref{loss for multiple}}

Let $\mathbf{M}=(m_{11},...,m_{1n_1},...,m_{N1}...,m_{Nn_N})^T$ be the vector of missing indicators. We use the first two terms of the Taylor expansion 
to approximate $Ch_{max}(\mathbf{D}_{\boldsymbol{\xi M}}\mathbf{R}^2(\boldsymbol{\xi}, \mathbf{M})\mathbf{D}_{\boldsymbol{\xi M}})$ at $\mathbf{M}=\mathbf{1}=(1,...,1)^T$. Notice that for $\mathbf{M}=\mathbf{1}$, $\mathbf{D}_{\boldsymbol{\xi 1}}$ is $\mathbf{D}_{\boldsymbol{\xi}}$ and $\mathbf{R}(\boldsymbol{\xi},\mathbf{1})=\mathbf{Z}(\mathbf{Z}^T\mathbf{D}_{\boldsymbol{\xi}}\mathbf{Z})^{-1}\mathbf{Z}^T:=\mathbf{R}(\boldsymbol{\xi})$. Then the approximation is 
\begin{eqnarray}
&&Ch_{max}(\mathbf{D}_{\boldsymbol{\xi M}}\mathbf{R}^2(\boldsymbol{\xi}, \mathbf{M})\mathbf{D}_{\boldsymbol{\xi M}}) \nonumber\\
&\approx & Ch_{max}(\mathbf{D}_{\boldsymbol{\xi}}\mathbf{R}^2(\boldsymbol{\xi})\mathbf{D}_{\boldsymbol{\xi}})+ \frac{\partial Ch_{max}(\mathbf{D}_{\boldsymbol{\xi M}}\mathbf{R}^2(\boldsymbol{\xi}, \mathbf{M})\mathbf{D}_{\boldsymbol{\xi M}}) }{\partial \mathbf{M}^T}|_{\mathbf{M}=\mathbf{1}} (\mathbf{M}-\mathbf{1}),
\nonumber
\end{eqnarray}
where 
$$ \frac{\partial Ch_{max}(\mathbf{D}_{\boldsymbol{\xi M}}\mathbf{R}^2(\boldsymbol{\xi}, \mathbf{M})\mathbf{D}_{\boldsymbol{\xi M}}) }{\partial \mathbf{M}}=\left(\begin{array}{c}
     \frac{\partial Ch_{max}(\mathbf{D}_{\boldsymbol{\xi M}}\mathbf{R}^2(\boldsymbol{\xi}, \mathbf{M})\mathbf{D}_{\boldsymbol{\xi M}}) }{\partial m_{11}}  \\
    ...\\
    \frac{\partial Ch_{max}(\mathbf{D}_{\boldsymbol{\xi M}}\mathbf{R}^2(\boldsymbol{\xi}, \mathbf{M})\mathbf{D}_{\boldsymbol{\xi M}}) }{\partial m_{Nn_N}}
\end{array}
\right)$$
and 
\begin{eqnarray}
&&\frac{\partial Ch_{max}(\mathbf{D}_{\boldsymbol{\xi M}}\mathbf{R}^2(\boldsymbol{\xi}, \mathbf{M})\mathbf{D}_{\boldsymbol{\xi M}}) }{m_{ij}} \nonumber\\
&=&\mathbf{v}_1^T\frac{\partial \mathbf{D}_{\boldsymbol{\xi M}} \mathbf{R}(\boldsymbol{\xi},\mathbf{M})}{\partial {m}_{ij}}\mathbf{R}(\boldsymbol{\xi},\mathbf{M})\mathbf{D}_{\boldsymbol{\xi M}} \mathbf{v}_1+ \mathbf{v}_1^T\mathbf{D}_{\boldsymbol{\xi M}} \mathbf{R}(\boldsymbol{\xi},\mathbf{M})\frac{\partial \mathbf{R}(\boldsymbol{\xi},\mathbf{M})\mathbf{D}_{\boldsymbol{\xi M}}}{\partial {m}_{ij}} \mathbf{v}_1,
\nonumber
\end{eqnarray}
 for $i\in \{1,...,N\}$ with $n\xi_i>0$, and $j=1,...,n_i$. 
Here $\mathbf{v}_1$ is the normalized eigenvector of $Ch_{max}(\mathbf{D}_{\boldsymbol{\xi M}}\mathbf{R}^2(\boldsymbol{\xi}, \mathbf{M})\mathbf{D}_{\boldsymbol{\xi M}})$ which is assumed to be simple (see Theorem 1 in \cite{Magnus1985}).

Next, we will find $\frac{\partial \mathbf{R}(\boldsymbol{\xi},\mathbf{M})\mathbf{D}_{\boldsymbol{\xi M}}}{\partial {m}_{ij}}$. We first notice that 
$$ \frac{\partial \mathbf{D}_{\boldsymbol{ \xi M}}}{\partial m_{ij}}=diag(\mathbf{e}_i)
$$
where $\mathbf{e}_i$ is an $N\times 1$ vector with the $ith$ element $e_i=1$ being the only nonzero element. 
Therefore,
\begin{eqnarray}
&&\frac{\partial \mathbf{R}(\boldsymbol{\xi},\mathbf{M})\mathbf{D}_{\boldsymbol{\xi M}}}{ \partial m_{ij}}=\mathbf{Z}\frac{\partial  (\mathbf{Z}^T\mathbf{D}_{\boldsymbol{\xi M}}\mathbf{Z})^{-1} }{\partial m_{ij}}\mathbf{Z}^T \mathbf{D}_{\boldsymbol{\xi M}} +\mathbf{R}(\boldsymbol{\xi},\mathbf{M}) \frac{\partial \mathbf{D}_{\boldsymbol{\xi M}}}{\partial m_{ij}}
\nonumber\\
&=& -\mathbf{R}(\boldsymbol{\xi},\mathbf{M})\frac{\partial \mathbf{D}_{\boldsymbol{\xi M}} }{\partial m_{ij}}\mathbf{R}(\boldsymbol{\xi},\mathbf{M}) \mathbf{D}_{\boldsymbol{\xi M}} +\mathbf{R}(\boldsymbol{\xi},\mathbf{M}) \frac{\partial \mathbf{D}_{\boldsymbol{\xi M}}}{\partial m_{ij}}
\nonumber\\ 
&=& - \mathbf{R}(\boldsymbol{\xi},\mathbf{M}) diag(\mathbf{e}_i) \mathbf{R}(\boldsymbol{\xi},\mathbf{M})\mathbf{D}_{\boldsymbol{\xi M}} + \mathbf{R}(\boldsymbol{\xi},\mathbf{M})diag(\mathbf{e}_i)
\nonumber\\
&=&   \mathbf{R}(\boldsymbol{\xi},\mathbf{M}) diag(\mathbf{e}_i)(\mathbf{I}- \mathbf{R}(\boldsymbol{\xi},\mathbf{M})\mathbf{D}_{\boldsymbol{\xi M}}).
\nonumber
\end{eqnarray}
We then have 
\begin{eqnarray}
&&\mathbf{v}_1^T\frac{\partial \mathbf{D}_{\boldsymbol{\xi M}}\mathbf{R}(\boldsymbol{\xi},\mathbf{M})}{\partial {m}_{ij}}\mathbf{R}(\boldsymbol{\xi},\mathbf{M})\mathbf{D}_{\boldsymbol{\xi M}} \mathbf{v}_1= \mathbf{v}_1^T\mathbf{D}_{\boldsymbol{\xi M}}\mathbf{R}(\boldsymbol{\xi},\mathbf{M}) \frac{\partial \mathbf{R}(\boldsymbol{\xi},\mathbf{M})\mathbf{D}_{\boldsymbol{\xi M}}}{\partial {m}_{ij}}\mathbf{v}_1,
\nonumber\\
&=& \mathbf{v}_1^T\mathbf{R}(\boldsymbol{\xi},\mathbf{M}) diag(\mathbf{e}_i)(\mathbf{I}- \mathbf{R}(\boldsymbol{\xi},\mathbf{M})\mathbf{D}_{\boldsymbol{\xi M}})\mathbf{v}_1,
\nonumber
\end{eqnarray}
and the first partial derivative of $Ch_{max}(\mathbf{D}_{\boldsymbol{\xi M}}\mathbf{R}^2(\boldsymbol{\xi}, \mathbf{M})\mathbf{D}_{\boldsymbol{\xi M}})$ with respect to ${m}_{ij}$ at $\mathbf{M}=\mathbf{1}$ is 
\begin{eqnarray}
2 \mathbf{v}_1^T\mathbf{R}(\boldsymbol{\xi})diag(\mathbf{e}_i)(\mathbf{I}- \mathbf{R}(\boldsymbol{\xi})\mathbf{D}_{\boldsymbol{\xi}})\mathbf{v}_1
=2\left( (\mathbf{I}- \mathbf{R}(\boldsymbol{\xi})\mathbf{D}_{\boldsymbol{\xi}})\mathbf{v}_1\mathbf{v}_1^T\mathbf{R}(\boldsymbol{\xi})\right)_{ii}.
\nonumber
\end{eqnarray}

For $\boldsymbol{tr}\left[\mathbf{Z}\left(\mathbf{Z}^T\mathbf{D}_{\boldsymbol{\xi}\mathbf{M}}\mathbf{Z} \right)^{-1}\mathbf{Z}^T\right] $, its first partial derivative  with respect to ${m}_{ij}$ at $\mathbf{M}=\mathbf{1}$ is 
\begin{eqnarray}
 -\boldsymbol{tr}\left[\mathbf{R}(\boldsymbol{\xi})diag(\mathbf{e}_i)\mathbf{R}(\boldsymbol{\xi}) \right]=-\left(\mathbf{R}^2(\boldsymbol{\xi})\right)_{ii}.
\nonumber
\end{eqnarray}

Then the first and second terms of the Taylor expansion of AMSE are
\begin{eqnarray}
&&\frac{\eta^2}{Nn} Ch_{max}(\mathbf{D}_{\boldsymbol{\xi}}\mathbf{R}^2(\boldsymbol{\xi})\mathbf{D}_{\boldsymbol{\xi}})+\frac{\sigma^2}{N}\boldsymbol{tr}\left[\mathbf{R}(\boldsymbol{\xi})\right]+\frac{\eta^2}{Nn}\nonumber\\
&&+ \frac{2\eta^2}{Nn}\sum_{i=1}^N\sum_{j=1}^{n\xi_i} (m_{ij}-1)\left( (\mathbf{I}- \mathbf{R}(\boldsymbol{\xi})\mathbf{D}_{\boldsymbol{\xi}})\mathbf{v}_1\mathbf{v}_1^T\mathbf{R}(\boldsymbol{\xi})\right)_{ii}\nonumber\\
&&+ \frac{\sigma^2}{N}\sum_{i=1}^N\sum_{j=1}^{n\xi_i} (1-m_{ij})\left(\mathbf{R}^2(\boldsymbol{\xi})\right)_{ii}. \label{MARE45}
\end{eqnarray}
Taking the expectation for $\mathbf{M}$ on (\ref{MARE45}) we then obtain an approximation of the design criterion MMPE. Notice that $E[m_{ij}]=p(\mathbf{x}_i,\boldsymbol{\gamma})$ which is the response probability at $\mathbf{x}_i$. 
Let $\mathbf{P}$ be the diagonal matrix of the vector  $(p(\mathbf{x}_i,\gamma))_{i=1}^N$. Then the approximated MMPE is
\begin{eqnarray*}
&&\frac{\eta^2}{Nn} Ch_{max}(\mathbf{D}_{\boldsymbol{\xi}}\mathbf{R}^2(\boldsymbol{\xi})\mathbf{D}_{\boldsymbol{\xi}})+\frac{\sigma^2}{N}\boldsymbol{tr}\left[\mathbf{R}(\boldsymbol{\xi})\right]+\frac{\eta^2}{Nn}\\
&&- \frac{2\eta^2}{Nn} \mathbf{v}_1^T\mathbf{R}(\boldsymbol{\xi})\mathbf{D}_{\boldsymbol{\xi}}(\mathbf{I}-\mathbf{P}) \mathbf{v}_1 +\frac{2\eta^2}{Nn} \mathbf{v}_1^T \mathbf{R}(\boldsymbol{\xi})\mathbf{D}_{\boldsymbol{\xi}}(\mathbf{I}-\mathbf{P}) \mathbf{R}(\boldsymbol{\xi})\mathbf{D}_{\boldsymbol{\xi}}\mathbf{v}_1\\
&& + \frac{\sigma^2}{N}\boldsymbol{tr}\left[ (\mathbf{I}-\mathbf{P})\mathbf{D}_{\boldsymbol{\xi}} \mathbf{R}^2(\boldsymbol{\xi})\right].
\end{eqnarray*}

\subsection*{Appendix E. Proof of Lemma \ref{asymaptotic distri}} 
Let $\mathbf{Y}=(Y_{11},...,Y_{1n_1},...,Y_{N1}, ...,Y_{Nn_N})$ where  $Y_{ij}$ is the $j$th observation being made at $\mathbf{x}_i$,  with $i=1,...,N$ and $j\leq n_j$. Let $m_{ij}$ be the missing indicator for $Y_{ij}$ as defined in  (\ref{MARE11}). The missing indicators satisfy $\sum_{i=1}^N\sum_{j=1}^{n_i}m_{ij}=O_p(n)$. Denote the maximum likelihood estimate (MLE) of the model coefficients as  $\hat{\boldsymbol{\beta}}$. Suppose that the missing data are handled by complete  case analysis. Then the MLE can be found by maximizing 
\begin{eqnarray}
-\frac{1}{2}\sum_{i=1}^N\sum_{j=1}^{n_i}m_{ij} \left[Y_{ij}-f(\boldsymbol{x}_i;\boldsymbol{\beta})\right]^2.
\nonumber
\end{eqnarray}
For convenience, we denote $-\frac{1}{2}\left[Y_{ij}-f(\boldsymbol{x}_i;\boldsymbol{\beta})\right]^2$ as $\Phi_{ij}(\boldsymbol{\beta})$. 
Next, we will find the asymptotic distribution of MLE conditional on the missing indicators $\mathbf{M}$. Notice that  MLE $\hat{\boldsymbol{\beta}}$ satisfies
\begin{eqnarray}
\sum_{i=1}^N\sum_{j=1}^{n_i}m_{ij}\dot{\Phi}_{ij}(\hat{\boldsymbol{\beta}}) = \mathbf{0}.
\nonumber
\end{eqnarray}
where
\begin{eqnarray}
\dot{\Phi}_{ij}({\boldsymbol{\beta}})
=\frac{\partial \Phi_{ij}({\boldsymbol{\beta}})}{\partial \boldsymbol{\beta}}= \left[Y_{ij}-f(\boldsymbol{x}_i;\boldsymbol{\beta})\right] \frac{\partial f(\boldsymbol{x}_i;\boldsymbol{\beta})}{\partial \boldsymbol{\beta}} .
\nonumber
\end{eqnarray}
Assume that $f^{(k)}(\boldsymbol{x};\boldsymbol{\beta})$ exists for $k=1,2, ...$. Then,  according to Taylor's Theorem, we have
\begin{eqnarray}
\mathbf{0}&=&\sum_{i=1}^N\sum_{j=1}^{n_i}m_{ij}\dot{\Phi}_{ij}(\hat{\boldsymbol{\beta}})\nonumber\\
&=&\sum_{i=1}^N\sum_{j=1}^{n_i}m_{ij}\dot{\Phi}_{ij}({\boldsymbol{\beta}})+\sum_{i=1}^N\sum_{j=1}^{n_i}m_{ij}\ddot{\Phi}_{ij}({\boldsymbol{\beta}})(\hat{\boldsymbol{\beta}}-{\boldsymbol{\beta}})
\label{MARE30}\\
&& +O(\| \hat{\boldsymbol{\beta}}-\boldsymbol{\beta} \|^2),
\nonumber
\end{eqnarray}
where 
\begin{eqnarray}
\ddot{\Phi}_{ij}({\boldsymbol{\beta}})=[Y_{ij}-f(\mathbf{x}_i;\boldsymbol{\beta})]\frac{\partial^2 f(\mathbf{x}_i;\boldsymbol{\beta})}{\partial \boldsymbol{\beta} \partial \boldsymbol{\beta}^T} -\frac{\partial f(\mathbf{x}_i;\boldsymbol{\beta})}{\partial \boldsymbol{\beta}} \left(\frac{\partial f(\mathbf{x}_i;\boldsymbol{\beta})}{\partial \boldsymbol{\beta} } \right)^T.
\end{eqnarray}
We then have
\begin{eqnarray}
\sqrt{n}(\hat{\boldsymbol{\beta}}-{\boldsymbol{\beta}})&=& \sqrt{n}\left(-\sum_{i=1}^N\sum_{j=1}^{n_i}m_{ij} \frac{\ddot{\Phi}_{ij}({\boldsymbol{\beta}})}{n} \right)^{-1}  \sum_{i=1}^N\sum_{j=1}^{n_i}m_{ij}\frac{ \dot{\Phi}_{ij}({\boldsymbol{\beta}})}{n} \nonumber\\
&&+ \left(-\sum_{i=1}^N\sum_{j=1}^{n_i}m_{ij} \frac{\ddot{\Phi}_{ij}({\boldsymbol{\beta}})}{n} \right)^{-1} \frac{O(\| \hat{\boldsymbol{\beta}}-\boldsymbol{\beta} \|^2)}{\sqrt{n}}.
\nonumber
\end{eqnarray}

Notice that since $\sum_{i=1}^N\sum_{j=1}^{n_i}m_{ij}\leq n $ we have 
\begin{eqnarray}
&&E_{\mathbf{Y}}\left[\sum_{i=1}^N\sum_{j=1}^{n_i}m_{ij} \frac{\ddot{\Phi}_{ij}({\boldsymbol{\beta}})}{n} \right]\nonumber\\
&=&\sum_{i=1}^N\sum_{j=1}^{n_i}m_{ij} \frac{E\left[ Y_{ij}-f(\mathbf{x}_i;\boldsymbol{\beta})  \right]}{n} \frac{\partial^2 f(\mathbf{x}_i;\boldsymbol{\beta})}{\partial \boldsymbol{\beta} \partial \boldsymbol{\beta}^T} -n^{-1} \sum_{i=1}^N\sum_{j=1}^{n_i}m_{ij} \frac{\partial f(\mathbf{x}_i;\boldsymbol{\beta})}{\partial \boldsymbol{\beta}} \left(\frac{\partial f(\mathbf{x}_i;\boldsymbol{\beta})}{\partial \boldsymbol{\beta} } \right)^T\nonumber\\
&=& n^{-1/2}\sum_{i=1}^N\sum_{j=1}^{n_i}m_{ij} \frac{\psi(\mathbf{x}_i;\boldsymbol{\beta})}{n}\frac{\partial^2 f(\mathbf{x}_i;\boldsymbol{\beta})}{\partial \boldsymbol{\beta} \partial \boldsymbol{\beta}^T}  -n^{-1}\sum_{i=1}^N\sum_{j=1}^{n_i}m_{ij} \frac{\partial f(\mathbf{x}_i;\boldsymbol{\beta})}{\partial \boldsymbol{\beta}} \left(\frac{\partial f(\mathbf{x}_i;\boldsymbol{\beta})}{\partial \boldsymbol{\beta} } \right)^T \nonumber\\
&=& O_p(n^{-1/2})-n^{-1} \sum_{i=1}^N\sum_{j=1}^{n_i}m_{ij} \frac{\partial f(\mathbf{x}_i;\boldsymbol{\beta})}{\partial \boldsymbol{\beta}} \left(\frac{\partial f(\mathbf{x}_i;\boldsymbol{\beta})}{\partial \boldsymbol{\beta} } \right)^T.\nonumber
\end{eqnarray}
The assumption  $\sum_{i=1}^N\sum_{j=1}^{n_i}(1-m_{ij})=O_p(1)$ implies that $$\frac{1}{n}\sum_{i=1}^N\sum_{j=1}^{n_i}m_{ij}\ddot{\Phi}_{ij}({\boldsymbol{\beta}})=\frac{1}{n}\sum_{i=1}^N\sum_{j=1}^{n_i}\ddot{\Phi}_{ij}({\boldsymbol{\beta}})-O_p(n^{-1}).$$ We then have
\begin{eqnarray}
-\sum_{i=1}^N\sum_{j=1}^{n_i}m_{ij}\frac{\ddot{\Phi}_{ij}({\boldsymbol{\beta}})}{n} -  n^{-1} \sum_{i=1}^N\sum_{j=1}^{n_i}m_{ij} \frac{\partial f(\mathbf{x}_i;\boldsymbol{\beta})}{\partial \boldsymbol{\beta}} \left(\frac{\partial f(\mathbf{x}_i;\boldsymbol{\beta})}{\partial \boldsymbol{\beta} } \right)^T\xrightarrow{pr} \mathbf{0}.
\nonumber
\end{eqnarray}

According to the above result, with $\mathbf{D}_{\boldsymbol{\xi}\mathbf{M}}=\hbox{diag}\left(\sum_{j=1}^{n\xi_1}m_{1j},..., \sum_{j=1}^{n\xi_N}m_{Nj}\right)$, we can find the mean and variance of  $\frac{1}{\sqrt{n}}\sum_{i=1}^N\sum_{j=1}^{n_i}m_{ij}\dot{\Phi}_{ij}({\boldsymbol{\beta}})$ conditional on $\mathbf{M}=(m_{11},..., m_{Nn_N})$ as follows:
\begin{eqnarray}
\frac{1}{\sqrt{n}}\sum_{i=1}^N\sum_{j=1}^{n_i}m_{ij} E_{\mathbf{Y}}\left[ \dot{\Phi}_{ij}({\boldsymbol{\beta}})\right]=\frac{1}{{n}}\sum_{i=1}^N\sum_{j=1}^{n_i}m_{ij} \psi(\mathbf{x}_i;\boldsymbol{\beta}) \frac{\partial f(\mathbf{x}_i;\boldsymbol{\beta})}{\partial \boldsymbol{\beta}} =\mathbf{Z}^T(\boldsymbol{\beta}) \mathbf{D}_{\boldsymbol{\xi M}} \boldsymbol{\Psi}(\boldsymbol{\beta})
\nonumber
\end{eqnarray}
and 
\begin{eqnarray}
&&Var_{\mathbf{Y}}\left[\frac{1}{\sqrt{n}}\sum_{i=1}^N\sum_{j=1}^{n_i}m_{ij}\dot{\Phi}_{ij}({\boldsymbol{\beta}}) \right]\nonumber\\
&=& Var_{\mathbf{Y}}\left[\sqrt{n}\sum_{i=1}^N\sum_{j=1}^{n_i}m_{ij}\left[Y_{ij}-f(\boldsymbol{x}_i;\boldsymbol{\beta})\right] \frac{\partial f(\boldsymbol{x}_i;\boldsymbol{\beta})}{\partial \boldsymbol{\beta}}  \right] \nonumber\\
&=&n Var_{\mathbf{Y}}\left[\sum_{i=1}^N\sum_{j=1}^{n_i}m_{ij}\left[Y_{ij}-f(\boldsymbol{x}_i;\boldsymbol{\beta})\right] \mathbf{z}_i(\boldsymbol{\beta}) \right]
\nonumber\\
&=&n Var_{\mathbf{Y}}\left[\mathbf{Z}^T(\boldsymbol{\beta}) \left(\begin{array}{c}
    \sum_{j=1}^{n\xi_1}m_{1j} (Y_{1j}-f(\boldsymbol{x}_1;\boldsymbol{\beta}))  \\
     ...\\
     \sum_{j=1}^{n\xi_N}m_{Nj} (Y_{Nj}-f(\boldsymbol{x}_N;\boldsymbol{\beta}) )
\end{array} \right) \right]
\nonumber\\
&=&n\sigma^2 \mathbf{Z}^T(\boldsymbol{\beta})\mathbf{D}_{\boldsymbol{\xi M}}\mathbf{Z}(\boldsymbol{\beta}),
\nonumber
\end{eqnarray}

By following the proof in  Section  12.2 of Seber and Wild (2003), we have  $\sqrt{n}(\hat{\boldsymbol{\beta}}-\boldsymbol{\beta})$ is asymptotic normal with asymptotic mean
$$(\mathbf{Z}^T(\boldsymbol{\beta})\mathbf{D}_{\boldsymbol{\xi M}}\mathbf{Z}(\boldsymbol{\beta}))^{-1}\mathbf{Z}^T(\boldsymbol{\beta}) \mathbf{D}_{\boldsymbol{\xi M}} \boldsymbol{\Psi}(\boldsymbol{\beta})
$$
 and asymptotic variance 
 \begin{eqnarray}
 && n\sigma^2(\mathbf{Z}^T(\boldsymbol{\beta})\mathbf{D}_{\boldsymbol{\xi M}}\mathbf{Z}(\boldsymbol{\beta}))^{-1}    \mathbf{Z}^T(\boldsymbol{\beta})\mathbf{D}_{\boldsymbol{\xi} M}\mathbf{Z}(\boldsymbol{\beta})   (\mathbf{Z}^T(\boldsymbol{\beta})\mathbf{D}_{\boldsymbol{\xi} M}\mathbf{Z}(\boldsymbol{\beta}))^{-1}\nonumber\\
 &=& n\sigma^2(\mathbf{Z}^T(\boldsymbol{\beta})\mathbf{D}_{\boldsymbol{\xi M}}\mathbf{Z}(\boldsymbol{\beta}))^{-1}.
 \nonumber
 \end{eqnarray}

Then, by the delta method, $\sqrt{n}(f(\mathbf{x}_i,\hat{\boldsymbol{\beta}})-f(\mathbf{x}_i,{\boldsymbol{\beta}}))$ follows an asymptotic normal distribution with asymptotic mean 
\begin{eqnarray}
\mathbf{z}_i^T(\boldsymbol{\beta})(\mathbf{Z}^T(\boldsymbol{\beta})\mathbf{D}_{\boldsymbol{\xi M}}\mathbf{Z}(\boldsymbol{\beta}))^{-1}\mathbf{Z}^T(\boldsymbol{\beta}) \mathbf{D}_{\boldsymbol{\xi M}}  \boldsymbol{\Psi}(\boldsymbol{\beta})
\end{eqnarray}
and asymptotic variance
\begin{eqnarray}
n\sigma^2 \mathbf{z}_i^T(\boldsymbol{\beta})(\mathbf{Z}^T(\boldsymbol{\beta})\mathbf{D}_{\boldsymbol{\xi M}}\mathbf{Z}(\boldsymbol{\beta}))^{-1}\mathbf{z}_i(\boldsymbol{\beta}).
\end{eqnarray}

\end{document}